\documentstyle[psfig,epsf,conf-X,10pt]{article}

\def\sub{{substructures}}
\def\apim{{$\Pi^{(m)}$}}
\def\api2{{$\Pi^{(2)}$}}
\def\api3{{$\Pi^{(3)}$}}
\def\api4{{$\Pi^{(4)}$}}
\def\h80{{$h_{80}^{-1}$}}
\def\L{{$\Lambda$CDM}}
\def\M{{CHDM}}
\def\gr{\kern 2pt\hbox{}^\circ{\kern -2pt K}} 

\def\ltsima{$\; \buildrel < \over \sim \;$}
\def\simlt{\lower.5ex\hbox{\ltsima}}
\def\gtsima{$\; \buildrel > \over \sim \;$}
\def\simgt{\lower.5ex\hbox{\gtsima}}
\begin{document} 
\small
\heading{%
%
Morphological evolution of X--ray clusters
using hydrodynamical simulations"
 
}
\par\medskip\noindent
\author{%
R.Valdarnini$^{1}$, S. Ghizzardi$^{2,4}$, S. Bonometto$^{3,4}$
}

\address{%
 SISSA Via Beirut 2-4, I34014, Trieste, Italy
}
\address{%
Istituto di Fisica Cosmica, CNR, via Bassini 15, I20133, 
Milano, Italy
}
\address{%
Dipartmento di Fisica dell' Universit\`a di Milano-BICOCCA,
    Via Celoria 16, I20133, Milano, Italy
}
\address{%
INFN -- Sezione di Milano, Italy
}

\begin{abstract}
A large set of TREESPH simulations is used to test
the global morphology of galaxy clusters and its evolution
against X--ray data. A powerful method to investigate
substructures in galaxy clusters are the
{\it power ratios} introduced by Buote $\&$ Tsai.
We consider three flat cosmological models: CDM, \L~($\Omega_\Lambda = 0.7$)
and \M~($\Omega_h = 0.2$, 1 massive $\nu$), all normalized
so to fit the observed number of clusters.
For each model we built 40 clusters, using
a TREESPH code, and performed a statistical comparison with a data 
sample including nearby clusters observed with ROSAT PSPC instrument.
The comparison disfavors the $\Lambda$CDM model,  as clusters
appear too relaxed, while CDM and CHDM clusters, in which a 
higher degree of complexity occurs, seem to be closer to observations.
A better fit of data can be expected for some different DM mix.
If DM distributions are used
instead of baryons, we find  substructures more pronounced 
than in gas and models have a different score. Using
hydrodynamical simulations is therefore essential to our aims.
\end{abstract}
\section{Introduction}

Clusters of galaxies are likely to be dynamically young systems and
a promising way to constrain cosmological models
arises from the study their substructures
(\cite{moh93}; \cite{bor93}).

Among the methods suggested to quantify the degree of inhomogeneity in 
clusters (\cite{moh95}; \cite{jin95}; \cite{ECF96} ), one of the most promising
is based on the so-called {\sl power ratios} \apim\ 
(\cite{VGB99}, hereafter VGB99; \cite{BT95}). It
amounts to a multipole expansion accounting for
the angle dependence of cluster X--ray surface brightness,
limited to the first few multipole terms and at fixed scales 
(typically of the order of the Mpc).
This is an effective and synthetic way to discriminate cluster features and
\apim\ are found do depend on the cosmological model 
enabling to discriminate among different cosmologies.

\section{Power ratios: definition and evolution}

The number of photons collected by ROSAT PSPC does not depend on the 
temperature $ T$, if it is$ \simgt 1$ keV.
We can then assume that the X--ray surface brightness 
$\Sigma_X = \Lambda \int{\rho_b^2 ({\bf r}) dz}$. Here $\Lambda \sim const$
results from an integration along the line of sight.

The procedure to work out \apim\ is as follows (see, e.g., \cite{G98}, 
hereafter G98; VGB99; \cite{BT95}):
(i) $\rho_b^2 ({\bf r})$ is projected along a line of sight
on a (random) plane to yield the $X$--ray surface 
brightness $\Sigma (R,\varphi)$; the centroid is used as origin. 
(ii) By solving the Poisson equation $\nabla^2 \Phi = \Sigma (R,\varphi) $
we obtain the pseudo--potential $\Phi(R,\varphi)$.
(iii) The coefficients of the expansion of $\Phi$ in plane harmonics 
will be used to build the power ratios 
$\Pi^{(m)} (R_{ap}) = \log_{10} (P_m/P_0) $. 
Here,
$P_m (R) = (\alpha_m^2 + \beta_m^2) /2 m^2 ~,
P_0 = [ \alpha_0 \ln(R/{\rm kpc}) ]^2$,
while
\begin{equation}
\alpha_m = \int_0^1 { ds\, s^{m+1} \int_0^{2\pi} { d\varphi [ \Sigma (sR,
\varphi) R^2] \cos(m\varphi) }},
\end{equation}
and $\beta_m$ has an identical definition, with sin instead of cos.
Owing to the definition of the centroid,
$\Pi^{(1)}$ vanishes. We shall restrict our analysis to 
 \apim ($m=2,3,4$), to account for \sub~ on scales not much below $R$ itself.
We consider three different aperture radii $R_{ap}=0.4,0.8,1.2 h^{-1}$Mpc.

Because of its evolution, a cluster moves along a curve of the 3--dimensional
space spanned by such \apim's; this curve is called {\sl evolutionary track}.
Quite in general, a
cluster starts from a configuration away from the origin, corresponding to a
large amount of internal structure and evolves towards isotropization and
homogeneization. 

Actual data, of course, do not follow the motion of a given cluster along the
evolutionary track. Different clusters, however, lie at different redshifts and
can be used to describe a succession of evolutionary stages.

\section{The simulated and the observed cluster sample}

We consider three spatially flat cosmological models:
CDM, \L\ with a cosmological constant accounting for $70\%$ 
of the critical density, and \M\ with 1 massive
neutrino with mass $m_\nu = 4.65\, $eV, yielding a HDM density parameter
$\Omega_h = 0.20$. We set 
$h=0.5$ for CDM and \M\ and $h=0.7$ for \L; for all models the primeval 
spectral index $n=1$ and the baryon density
parameter is selected to give $\Omega_b h^2 = 0.015$.  
All models were normalized in order to reproduce the present
observed cluster abundance (\cite{ECF96}, \cite{gir98}). 

In order to achieve a safe statistical basis for our analysis, for 
each cosmological model,
we select the 40 most massive clusters from an N--body P3M 
simulation.
For each of them, we  perform a hydrodynamical TREESPH simulation 
(VGB99, G98; see also \cite{HK89} and  \cite{NFW95}).

Clusters are distributed in redshift so to reproduce the same 
redshfit distribution of the observed cluster sample.

Our observed data set is the same used by \cite{BT96}, 
including nearby 
($z \simlt 0.2$) clusters observed with ROSAT PSPC instrument 
(see VGB99, G98 for details).
The resulting sample is partially incomplete, but, clusters 
were not selected for reasons related to their morphology and the missing 
clusters are expected to have a distribution of power ratios similar to the 
observed one.

\section{Results and Conclusions}

For simulated clusters, power ratios \apim\ have been computed 
from the gas distribution.

A visual inspection of how the \apim\ are distributed can be obtained
from Fig.1, whose histograms show the 
fraction of clusters with a given \apim\ . 
Here we show the distribution  
for $\Pi^3$ and $R_{ap}=0.8 h^{-1}$Mpc for each cosmological model and for 
the ROSAT data sample; distributions for
the other \apim\ and the other apertures show a similar behavior

\begin{figure*}
\centerline{\mbox{\epsfxsize=10.0truecm\epsffile{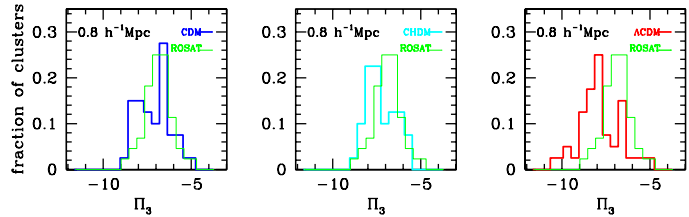}}}
\caption[]{Histograms of $\Pi^3$ in the $R_{ap}=0.8 h^{-1}$Mpc for simulated 
clusters.}
\end{figure*}

Quite in general we can conclude that while CDM and \M~
are marginally consistent with data, \L~ is far below them. 
In order to quantify these differences, we used  
the Student t--test, the F--test and the Kolgomorov--Smirnov (KS) test. 
For example, according to the t--test, 
the probability $p$--$t$ that the simulated and 
observed power ratios distributions are originated from the same process 
is roughly in the ranges 0.11--0.60 for CDM, 
0.03--0.37  for \M and  
0.15$\cdot 10^{-4}$--0.87$\cdot 10^{-2}$ for \L.
The other statistical tests provide similar probabilities.
Such figures seem to exclude that \L\ can be considered a reasonable
approximation to data. The best score belongs to CDM, but also \M\ is not fully
excluded and different mixtures could certainly have better performance. 
An inspection of the
model clusters actually shows that the \L~ model does produce less
substructures than the other models do. A possible interpretation
of such output is that the actual amount of substructures is  governed by
$\Omega_0$ rather than by the shape of power spectra. 

According to the same tests, if cosmological models are compared with 
data on the basis of DM \apim, values are shifted, indicating an 
increase in the amount of substructures for DM with respect to the gas.
This is to be ascribed to the smoothing 
effects of the interactions among gas particles, which erase anisotropies and 
structures, while DM \apim scarcely feel dissipative processes.
Hence, using DM \apim leads to biased scores:
CDM and \M\ models keep too many substructures and are 
no longer consistent with data;
on the contrary, the increase of substructures pushes \L\ to agree
with ROSAT sample outputs.

We also considered the cluster distribution in the 3--dimensional 
parameter space 
with axes given by \apim\ ($m=2,3,4$), as well as projections of such 
distributions on planes.
Comparing such distributions for data and models, we find a significantly
stronger correlation of \apim\ in models than in data. 
Distributions for simulated clusters 
show a linear trend while distributions 
of observed clusters tend to be more scattered than simulated points.
The degree of 
correlation depends on the model, but seems however in disagreement with data.
Model clusters tend to indicate a significantly faster evolution than data. The
cosmological model which seems closest to data is \M\ and it is possible that
different \M\ mixtures can lead to further improvements.
Also $\Lambda$ models with $\Omega_m > 0.5$ might deserve to be
explored. Virialized clusters had their turn--around at a time
$< t_o/3$ ($t_o$: present age of the Universe). In turn, $\Omega_\Lambda$
becomes dominant at $z = (\Omega_\Lambda/\Omega_m)^{1/3}$. If
such redshift occurs at a time $\sim t_o/3$, we expect results
from $\Lambda $ models to be closer to observations.

\vfill
\end{document}